# Quantitative Measurement of Heritability in the Pre-RNA World

Norichika Ogata[1,2]
[1]Medicale Meccanica Inc., Kawasaki, Japan,
[2]Nihon BioData Corporation, Kawasaki, Japan
norichik@nbiodata.com

*Abstract*— Long ago, life obtained nucleotides in the course of evolution and became a vehicle for them[1]. Before assembly with nucleotides, in the pre-RNA era, what system dominated heredity? What was the subject of survival competition? Is it still a subject of competition? Self-organized complex systems are hypothesized to be a primary factor of the origin of life and to dominate heritability[2,3], mediating the partitioning of an equal distribution of structures and molecules at cell division[4]. The degree of strength of self-organization would correlate with heritability; self-organization is known to be a physical basis of hysteresis phenomena[5], and the degree of hysteresis is quantifiable[6]. However, there is no argument corroborating the relationship between heritability and hysteresis. Here, we show that the degree of cellular hysteresis indicates its heritability and daughter equivalence at cell division. We found a correlation between thermal hysteresis in cell size and heritability, which quantified cell line generation stability, suggesting that the self-organized complex system is a subject of survival competition, comparable to nucleotides[7]. Furthermore, in single-cell-resolution observations, we found that thermal hysteresis in cell size indicates equivalent partitioning in future cell division. Our results demonstrate that self-organized complex systems contribute to heredity and are still important in mammalian cells. Predicting the cell line generational stability required for the industrial production of therapeutic biologics[8] is useful. Discovering ancient and hidden heredity systems enables us to study our own origin, to predict cell features and to manage them in the bio-economy.

## I. Result and Discussion

The existence of quasi-periodic crystals as hereditary information storage, as predicted by Schrödinger, was corroborated by the discovery of DNA [7]. This phenomenon has allowed the scientists pioneering enumerative molecular biology and technological developments in DNA sequencing to discover genomes every day. Even so, we have not been able to predict the future beyond a few months for cultured cells, the simplest isolated life, which is widely used for therapeutic protein productions and cellular biological experiments. A well-known unwanted cellular transformation is reduction in therapeutic protein production, and it is unpredictable which cell strain will be transformed[8]. An explainable phenomenon should be able to predict this future development. A quasi-periodic crystal would not be sufficient to explain hereditary phenomena.

Heredity is fundamentally an issue of information and hysteresis, fundamentally. Quasi-periodic crystals such as RNA and DNA are hereditary information storages because of their hysteretic behaviour. Other crystals and self-organized soft matter systems, such as liquid crystals, colloids, block copolymers[9], catalytic networks[10] and liquid-liquid phase separations[11] are also regarded as potential information storage

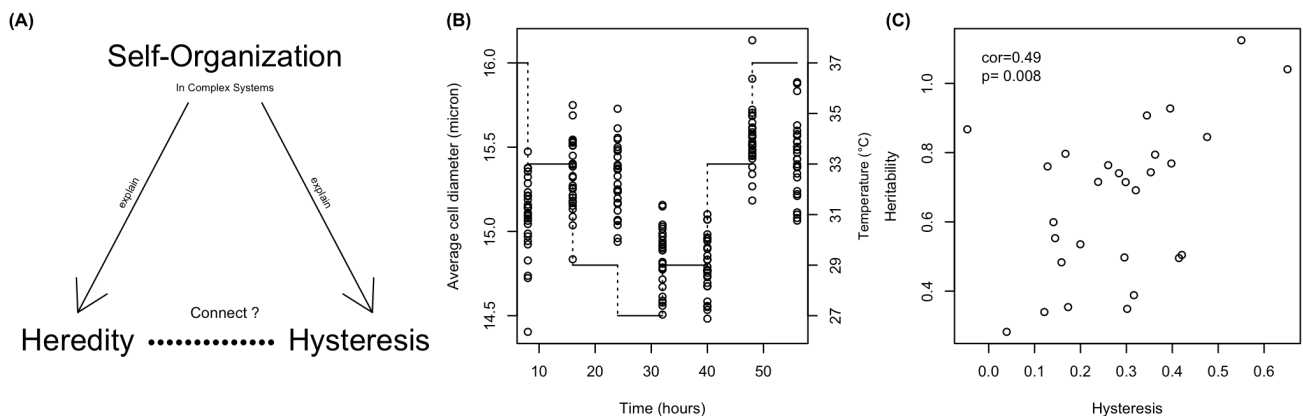

**Fig 1. A conceptual connection between hysteresis and heredity, the sons of self-organization in complex systems.**

(A) An illustration of the conceptual connections. (B) Thermal hysteresis of the cell size of 28 clonal cell strains. Culture time and cell size were plotted. (C) Comparison between hysteresis and heritability. Hysteretic loop size estimated by balance of cell sizes in 29 °C and heritability estimated by the ratio of protein production in cell strains initially and after 15 generations were plotted.

since self-organization is the physical basis of hysteresis phenomena[5,12,13]. These self-organized complex systems are thought to be not only the origin of life and heredity[2,3,10] but also the driving force partitioning cellular structures and molecules in equal distribution at division[4]. Equal partitioning in cell division ensures similarity between parent cells and daughter cells; this similarity across generations is the principle nature of heredity. As described above, two conceptual connections between self-organization and heredity and between self-organization and hysteresis have been explained theoretically and experimentally. Here, we hypothesized that there is some relationship between heredity and hysteresis (Figure 1a). To test our hypothesis, we should quantify and compare heritability and hysteresis.

Measuring heritability is a common task in therapeutic protein production, in the form of cell line generation stability[8]. In contrast, measuring hysteresis is rare. There are several hysteretic phenomena in biology[14], but the cellular hysteretic phenomenon is known in few cases[15,16]. Here, we measured cell size thermal hysteresis and the temperature-dependent volume phase transition behaviour since several thermal hysteretic phenomena were previously studied in several soft materials[17,18], and a hint was provided by a reported non-linear thermal effect on cell size[19]. To survey the cell size thermal hysteresis, we observed 28 sub-cloned CHO cell strains cultured in an iterative thermal changing environment. The cells were suspended in an environment with changing temperature: 37 °C (over 24 hours), 33 °C (8 hours), 29 °C (8 hours), 27 °C (8 hours), 29 °C (8 hours), 33 °C (8 hours) and 37 °C (8 hours). Every 8 hours, the cellular sizes of each cell strain were measured using the Vi-Cell cell counter. The cell size thermal hysteresis was observed, and it was revealed that we can measure the size of hysteretic loops in differences between cell size in the first 29 °C and second 29 °C period for each cell strain. In contrast, thermal hysteresis was not observed in transcriptomes[16,20] (Supplemental Figure 1), indicating that this thermal hysteresis is independent of genome expression.

After measuring the hysteretic characteristics of 28 new sub-cloned protein (mAb)-producing CHO cell strains in the same manner, the cell strains were sub-cultured 15 times, and a generation-dependent decrease in protein production was measured as an indicator of heritability. The results showed that the sizes of the hysteretic loops of each cell strain were correlated with the degree of heritability (Figure 1c). The differences in heritability defined in this study between cell strains indicated that self-organized complex systems are the subject of survival competition in the pre-RNA world and today (Supplemental Figure 2).

For a deeper insight into this heritability, we observed adherent cultured cells in the iterative temperature changing environment, taking pictures at 1 minute intervals (https://youtu.be/PNcIVrXgyA8). In total, 106 cells were observed for 40 hours: 19 cells did not divide, 61 cells divided once (2 daughter cells), 9 cells divided twice (3 daughter cells), 9 cells divided 3 times (4 daughter cells), 3 cells moved out of the field of view, 4 cells re-fused after division, and one cell divided into 3 daughter cells at once (Supplemental Figure 3,

4). The cell size thermal hysteresis could be analysed only in 6 cells, which divided after the 2nd 33 °C period. Three cells, whose cellular size at 27 °C was smaller than the average size at 29 °C, showed equal cell division, and the other three cells, whose cellular size at 27 °C was larger than the average size at 29 °C, showed non-equal cell division. In these cells, the sizes of the daughter cells were different, and the pseudopodium formation timing differed in each cell (Supplemental Figure 5). Our observation indicated again that the concept of "clonal cell" is not well defined[21,22]. Several cells showed fusion, and the cell that was divided into 3 daughter cells at once.

A hidden connection between two phenomena explained in the context of the self-organization of complex systems was revealed in this study. Further experimental and theoretical corroborations of the connections between various phenomena that can be explained by the self-organization of complex systems are needed. We are not the only potential users of the connections between self-organization-related phenomena; cells in non-artificial environments might also be such users. The cells of multicellular organisms might use iterative environmental changes to watch bystander cells, for example, establishing the circadian rhythm, which was not observed in non-photosynthetic bacteria, yeast and embryos immediately after fertilization.

## II. MATERIALS AND METHODS

Cell strains and cultures

The monoclonal protein (mAb)-producing CHO S cell strains were obtained from the CHO S cell line (Gibco, Paisley, UK) as previously described [23]. The heavy chain (HC) of human IgG1 was inserted into EcoRI- and NheI-digested pEHX1.1 (Toyobo) (pEHC), and the light chain (LC) of human IgG1 was inserted into EcoRI- and NheI-digested pELX2.1 (Toyobo, Osaka, Japan) (pELC). The internal ribosome entry site (IRES) fragment obtained from a pIRES plasmid (Takara Bio Inc., Shiga, Japan) and the dhfr fragment obtained from a pSV2-dhfr/hGM-CSF plasmid were inserted into MluI- and NotI-digested pEHC (pEHC-IRES-dhfr). The pEHC-IRES-dhfr and pELC plasmids were digested with EcoRI and BglII and integrated into one plasmid. The antibody expression plasmid linearized by SspI was transfected into CHO S cells using X-tremeGENE 9 DNA Transfection Reagent (Roche Applied Science, Mannheim, Germany). Stable cells were selected in medium containing 15 mg/mL puromycin (InvivoGen, San Diego, CA, USA). Individual clones were isolated from the stable cell pool by limiting dilution. A total of 56 isolated clones were used in this study.

The CHO S cells were cultured in suspension using T-25 flasks (Grainer, Nürtingen, Germany) on a rotary shaker (120 rpm) (TAITEC, Saitama, Japan) with 7 mL of BalanCD CHO Growth A media (Irvine Scientific, CA, USA). The suspension culture condition was assessed using computational fluid and particle dynamics analyses with OpenFOAM[24] version 2.3.0. OpenFOAM was installed on the Ubuntu version 14.04 operating system (http://www.ubuntu.com) on a MacBook Pro (13-inch, early 2011; 16 GB, 1333 MHz DDR3; Apple, CA, USA). We modified stirringInterPTFoam to compute the culture conditions for box-shaped T-25 flasks on a rotary shaker (https://github.com/akionux/OpenFOAM-2.3.x/tree/master/applications/solvers/multiphase/stirringInterFoam/stirringInterPTFoam). All cells were cultivated in a humidified atmosphere containing 5% $CO_2$.

The CHO K1 cells were maintained in Dulbecco's Modified Eagle Medium: Nutrient Mixture F-12 (DMEM/F-12) (Gibco, Paisley, UK) with 10% dialysed foetal bovine serum (FBS) (SAFC Biosciences, Lenexa, KS).

For the examination of hysteresis, the culture temperature was changed iteratively: 37 °C (over 24 hours), 33 °C (8 hours), 29 °C (8 hours), 27 °C (8 hours), 29 °C (8 hours), 33 °C (8 hours) and 37 °C (8 hours).

Cell culture observation

Cellular diameters and viable and total cell concentrations were determined using a Vi-cell XR (Beckman Coulter, Fullerton, CA, USA). Adherent cell culture was performed on a BioStudio-T (Nikon, Kanagawa, Japan).

Evaluation of antibody productivity

Recombinant IgG antibody concentration was quantified with a Cedex Bio instrument (Roche Diagnostics GmbH, Germany).

Statistical analysis

The correlation between hysteresis and heritability was statistically tested using R (version 3.5.1) and the cor.test function (Pearson's product-moment correlation) [25].

Library preparation, sequencing and alignment

We prepared a library of CHO K1 in the iterative thermal changing environment for RNA-seq using a commercial kit (TruSeq RNA Sample Kit; Illumina, San Diego, CA, USA) in accordance with the manufacturer's protocol. We sequenced libraries for conventional RNA-seq using a commercial sequencer (NextSeq 500; Illumina) in accordance with the manufacturer's protocol. Short-read data have been deposited in the DNA Data Bank of Japan (DDBJ)'s Short Read Archive under project ID DRA007812. All short reads were mapped to CHO-K1 RefSeq assembly (22,516 sequences, RefSeq Assembly ID: GCF_000223135.1) and CHO-K1 mitochondrial DNA (1 sequence, RefSeq Assembly ID: GCF_000055695.1) using bowtie2 (version 2.3.0) [26]. Gene and isoform expression levels were estimated from RNA-seq data using RSEM (version 1.2.31) [27]. The Kolmogorov complexity was estimated from RNA-seq data as previously described[20].

Supplemental Figure 1

Comparison between the Kolmogorov complexity of the transcriptome and iterative environmental changes.

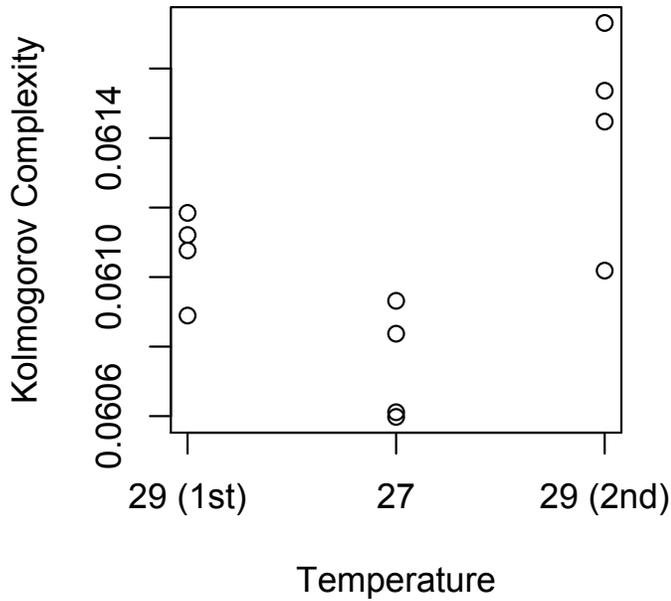

Supplemental Figure 2

Thermal hysteresis of cell size in each cell strain. Temperature and cell size are plotted. The upper 4 cell strains have strong heritability, and the lower 4 cell strains have weak heritability.

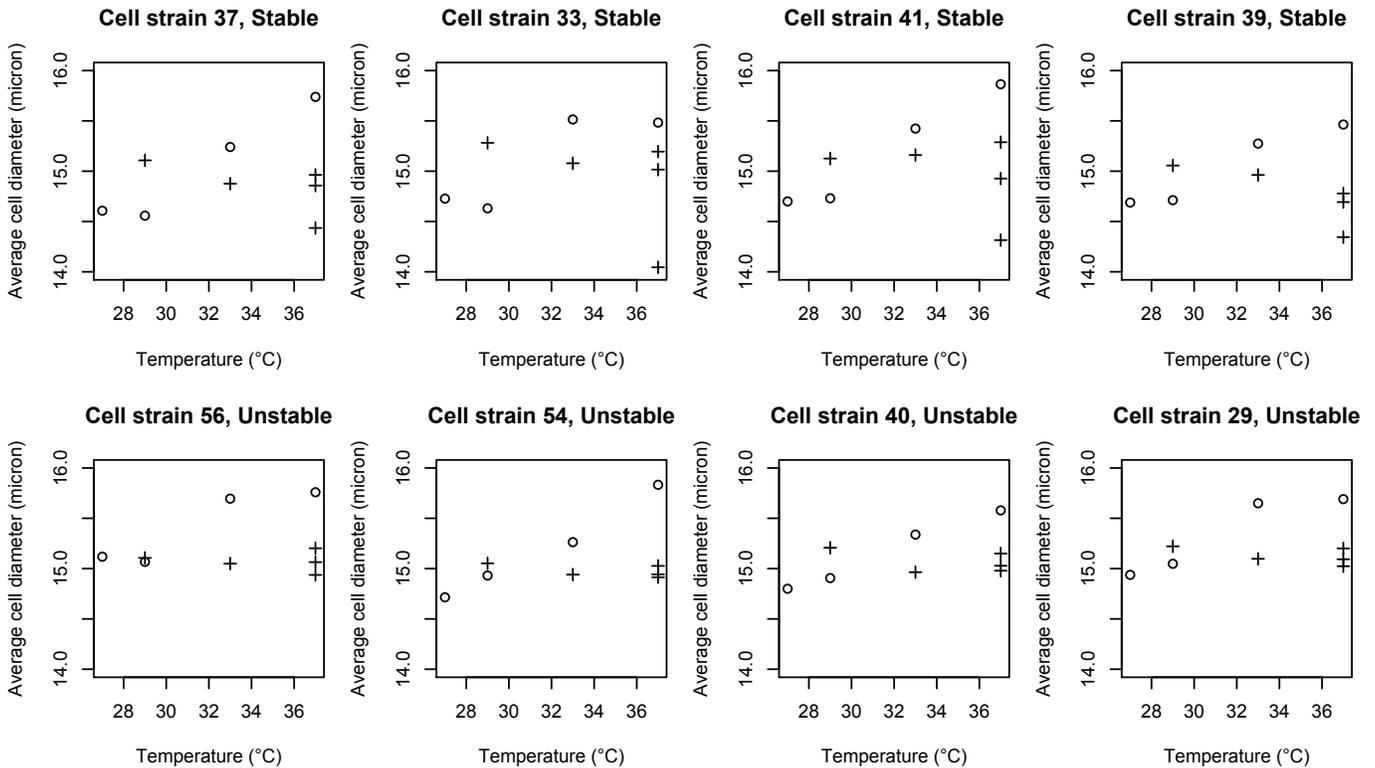

Supplemental Figure 3

Cells identified on a BioStudio-T. In total, 106 cells were observed for 40 hours: 19 cells did not divide (7, 10, 15, 18, 24, 35, 38, 43, 44, 49, 51, 63, 67, 71, 75, 85, 86, 89, 100); 61 cells divided once (2 daughter cells, cell ID: 1, 2, 3, 8, 11, 12, 13, 16, 19, 20, 21, 22, 23, 26, 27, 28, 29, 30, 31, 32, 34, 36, 37, 39, 40, 41, 42, 45, 46, 52, 53, 54, 55, 56, 58, 59, 61, 64, 65, 66, 70, 73, 74, 76, 79, 81, 82, 83, 84, 88, 90, 91, 92, 93, 95, 96, 97, 98, 99, 103, 104, 106); 9 cells divided twice (3 daughter cells, cell ID: 6, 9, 48, 50, 68, 69, 87, 94, 102); 9 cells divided 3 times (4 daughter cells, cell ID: 14, 25, 33, 60, 62, 78, 80, 84, 105) 3 cells moved out of the field of view (cell ID: 4, 57, 101) 4 cells re-fused after division (cell ID: 5, 47, 72, 77); and a single cell divided into 3 daughter cells at once (cell ID: 17).

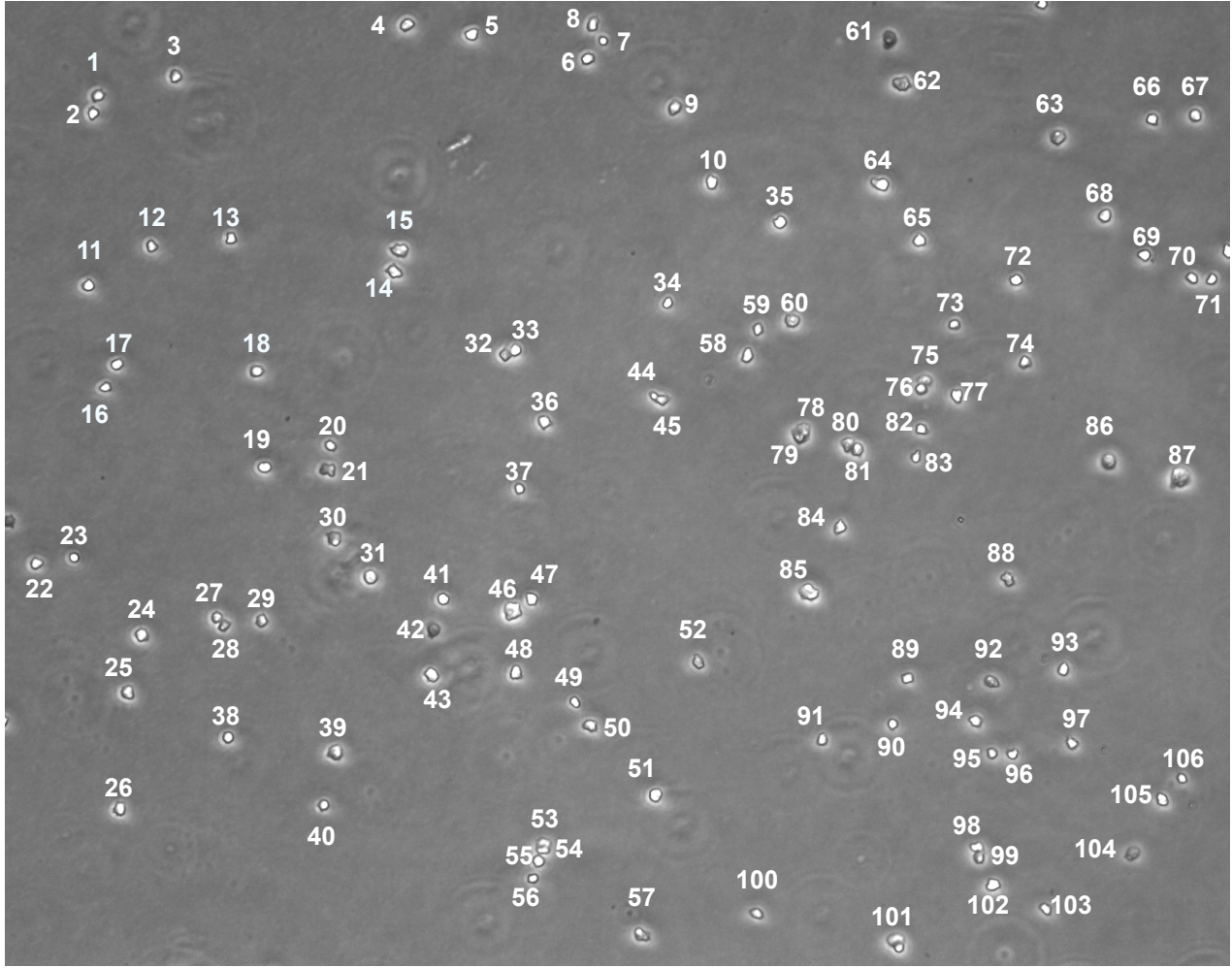

Supplemental Figure 4
Cell divisions in 98 cells. Each line indicates a cell, and the line changes colour when the cell divides.

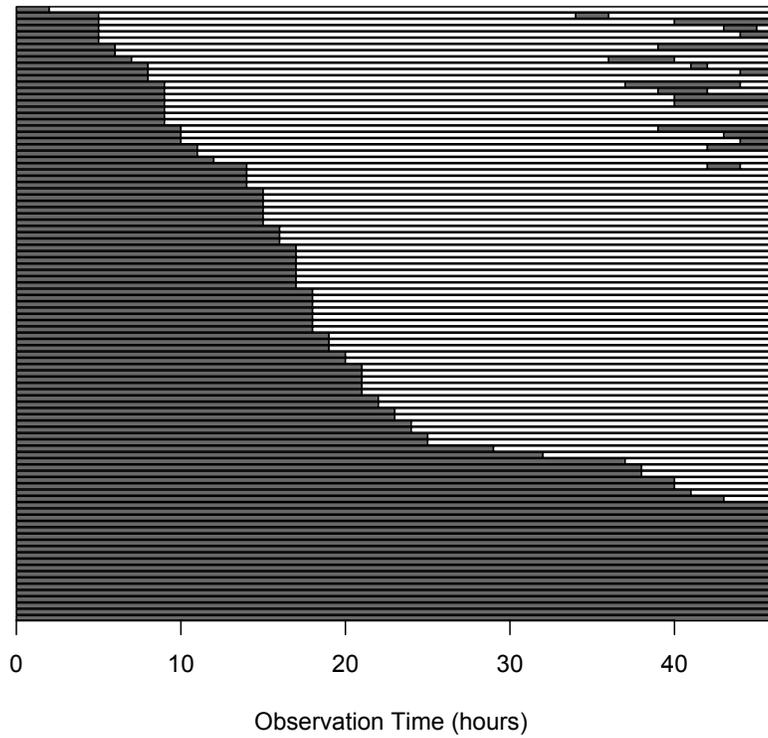

Observation Time (hours)

Supplemental Figure 5
Single-cell hysteresis and heredity. Cell sizes were measured at 29 °C, 27 °C and 29 °C, and their equality of cell division was evaluated.

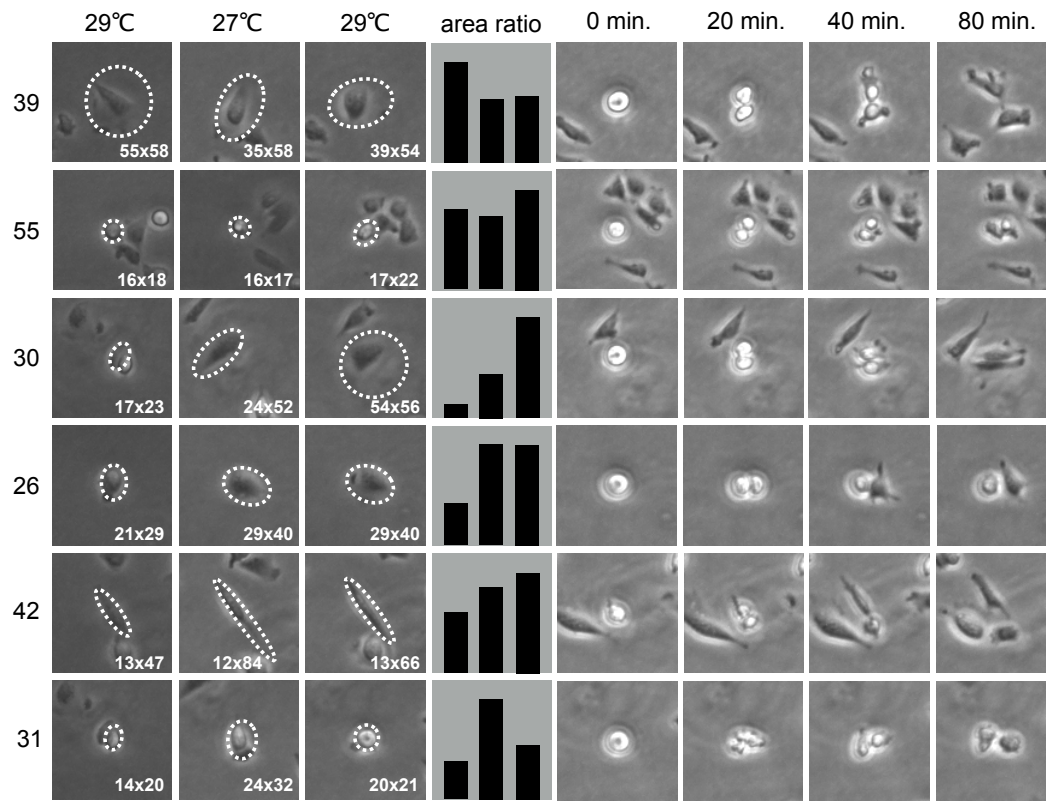